\documentclass[review]{elsarticle}

\usepackage{hyperref}
\usepackage{url}
\usepackage[linesnumbered,lined,boxed,commentsnumbered]{algorithm2e}
\usepackage{graphicx,subcaption}
\usepackage{caption}
\usepackage{capt-of}
\usepackage{lscape}
\usepackage{rotating}
\usepackage{appendix}
\usepackage{amssymb}
\usepackage{xcolor}
\usepackage{multirow}

\journal{Journal of Computer Communications}

\begin{document}

\begin{frontmatter}

\title{OLCPM: An Online Framework for Detecting Overlapping Communities in Dynamic Social Networks}

\author[add1]{Sou\^{a}ad Boudebza
\corref{mycorrespondingauthor}}
\cortext[mycorrespondingauthor]{Corresponding author}
\ead{s\_boudebza@esi.dz}

\author[add3,add5]{R\'{e}my Cazabet}
\author[add1]{Fai\c{c}al Azouaou}
\author[add4]{Omar Nouali}

\address[add1]{Ecole nationale Sup\'{e}rieure d'Informatique, BP 68M, 16309, Oued-Smar, Alger, Alg\'{e}rie. http://www.esi.dz}
\address[add3]{Sorbonne University, UPMC University, CNRS, LIP6 UMR 7606, Paris, France.}
\address[add4]{Division de Recherche en Théorie et Ingénierie des Systèmes Informatiques,CERIST, Rue des Frères Aissiou, Ben Aknoun, Alger, Algérie. }
\address[add5]{Univ Lyon, Universite Lyon 1, CNRS, LIRIS UMR5205, F-69622 France.}

\begin{abstract}
Community structure is one of the most prominent features of complex networks. Community structure detection is of great importance to provide insights into the network structure and functionalities. Most proposals focus on static networks. However, finding communities in a dynamic network is even more challenging, especially when communities overlap with each other. In this article, we present an online algorithm, called OLCPM, based on clique percolation and label propagation methods. OLCPM can detect overlapping communities and works on temporal networks with a fine granularity. By locally updating the community structure, OLCPM delivers significant improvement in running time compared with previous clique percolation techniques. The experimental results on both synthetic and real-world networks illustrate the effectiveness of the method.
\end{abstract}


\end{frontmatter}

\section{Introduction}


The analysis of complex networks is a fast growing topic of interest, with applications in fields as various as neural networks, protein networks, computer networks or geographical networks. One of the most prominent application domain is social network analysis.

The study of social networks can be traced back to the beginning of the 19th century, since the initial work on sociometry \protect\cite{moreno1934}. This subject has gained new momentum in recent years, mainly due to the advent of the information age and internet, which has led to the extensive popularity of online social networks, producing large social datasets that can be studied by researchers. The goal of social network analysis is to analyze relationships among social entities and to understand the general properties and features of the whole network, typically by means of graph theory. Nodes in the graph represent social actors within the network (people, organization, groups, or any other entity) and edges characterize social interactions or relations between nodes (friendship, collaboration, influence, idea, etc.). 

One of the most prominent features of social networks is their community structure, characterized by the existence of nodes collections called communities, where nodes within a collection tend to interact more with each other than with the rest of the network \cite{radicchi2004}. Individuals within the same community often share similar properties, such as interests, social ties, location, occupation, etc. Therefore, the ability to detect such communities could be of utmost importance in a number of research areas, such as recommender systems \cite{boratto2009}\cite{deng2014}, email communication \cite{moradi2012}, epidemiology \cite{kitchovitch2011}, criminology \cite{ferrara2014}, marketing and advertising \cite{mckenzie1999, fenn2009}, etc.

There are many challenges facing community detection. One of the most important, in particular for social networks, is \textit{overlap of communities}: in such networks, individuals often belong to several social groups. For instance, individuals often belong to familial and professional circles; scientists collaborate with several research groups, etc. The second challenge lies in the fact that real-world communities are time-evolving. The community structure changes as the social entities and their interactions evolve. These changes can be modeled as addition and removal of nodes and edges from the graph. For instance, in online social networks like Facebook, changes are introduced by users joining or withdrawing from the network, or by people adding each other as "friend". These changes may lead to a significant transformation of the network community structure. Palla et al.\cite{palla2007} propose six types of events which may occur during the evolution of communities: birth, growth, shrink, merge, split, and death. The communities can grow or shrink, as members are added or removed from an existing community. As time goes by, new communities can be born, and old communities may disappear. Two communities can become closely related and merge into a single one, or, conversely, a single community can split into two or more distinct ones.

\subsection{Rationale for an online version of the Clique Percolation Method}
A growing number of methods have been proposed to reveal overlapping and evolving community structures \cite{wang2013communities,cazabet2014dynamic}. One of the most prominent of these methods was proposed by Palla et al.\cite{palla2007}. The clique percolation method (CPM) \cite{palla2005} is used to extract the community structure at each time step of an evolving network. Then, communities in consecutive time steps are matched. 

The CPM method, thanks to its community definition, has interesting properties compared with other popular methods such as Louvain and infomap \cite{louvain,rosvall2008maps}: 
\begin{itemize}
\item It is deterministic, i.e., two runs on two networks with the same topology will yield the same results.
\item Communities are defined intrinsically, i.e., each community exists independently from the rest of the network, unlike methods using a global quality function such as the \textit{modularity} \cite{girvan2002}, that suffer from resolution limits \cite{fortunato2007resolution} binding the size of communities to the size of the network.
\item Communities can overlap, i.e., a node can be part of several communities.
\end{itemize}

These properties represent an advantage when working with social networks and with dynamic networks. In particular, a well-known problem with the discovery of evolving communities is the so-called instability of methods \cite{aynaud2010static}, which can be summarized as follows: because community detection methods are unstable, the difference observed in the partition between two consecutive periods of the network might be due either to significant changes in the network or to random perturbations introduced by the algorithm itself. This problem is due to (1) the usage of stochastic methods, as two runs on very similar (or even identical) networks can yield very different results if the algorithm reaches different local maximum, (2) non-intrinsically defined communities, as a modification of a community might be due to changes introduced in an unrelated part of the network. 

Given these observations, CPM appears as a natural candidate to be used for dynamic community detection. The method adapting CPM to the dynamic case \citep{palla2007}, however, suffers from at least two weaknesses for which we propose solutions in this article, one due to CPM itself, and other to its adaptation to the dynamic case:
\begin{itemize}
\item All cliques need to be discovered anew at each step, both in the new graph snapshot and in a joint graph between snapshots at $t$ and $t-1$, which is computationally expensive for networks with many steps of evolution.
\item Nodes must belong to a cliques of size at least $k$ to be part of a community, and as a consequence, some nodes might not be affected to any community. As most social networks have a scale-free degree distribution, a large number of nodes remain without a community.
\end{itemize}

To circumvent these issues, we propose a new two-step framework for detecting overlapping and evolving communities in social networks. First, built upon the classical algorithm CPM, we introduce an Online CPM algorithm (OCPM) to identify the core nodes of communities in real time. To do that, we propose to use \textit{stream graph} as a network model. At every change in the network, the community structure is updated at the local scale. This allows significant improvements in computational complexity compared with dynamic CPM \cite{palla2007}. Second, to deal with the coverage problem of CPM, we propose a label propagation post-process (OLCPM)and thus, nodes not embedded in any community will be assigned to one or more communities.

The rest of the paper is organized as follows: section \ref{relatedWork} discusses the related work on overlapping and evolving community detection algorithms. 
In Section \ref{dynamicNetworkModel}, we present the different types of dynamic networks and introduce a fully dynamic network model. 
Section \ref{OLCPMsection} presents the OLCPM framework of dynamic community detection: OCPM algorithm and Label propagation based post process. 
Experimental results are described in section \ref{Experiments}.

\section{Related work}
\label{relatedWork}
In this section, we first introduce the Clique Percolation Method (CPM) \cite{palla2005} and its dynamic version \cite{palla2007}, on which our proposal is built on. Then, we present a brief overview of some relevant research work on overlapping and dynamic community detection.  

Palla et al.\cite{palla2007} were among the first to propose an approach for dealing with dynamic and overlapping community detection. Their approach has two main steps: i) static community identification and ii) community matching. In the first step, the CPM method \cite{palla2005} is used to extract the community structure at each time step. In this method, a community is defined as the union of all \textit{$k$-cliques} (complete subgraphs of size $k$) that can be reached from each other through a series of adjacent $k-cliques$ (sharing $k-1$ nodes). In the second step, communities are matched between consecutive snapshots. The following process is used: for each pair of consecutive snapshots, a joint graph is created, containing the union of nodes and links from both networks. CPM is then applied to the resulting graph. The communities in the joint graph provide a natural connection between communities in the consecutive snapshots. If a community in the joint graph contains a single community in each corresponding snapshot, then they are matched. If the joint graph contains more than one community from either snapshot, the communities are matched in descending order of their relative node overlap. Overlap is computed for every pair of communities from the two snapshots as the fraction of the number of common nodes to the sum of the number of nodes in both communities. 

The work of Palla et al. \cite{palla2007} falls into the category of community matching approaches, i.e., methods with a static community detection step and a matching step. Most of the earliest algorithms proposed for dynamic community detection were following a similar approach, with variations in the method used for detection in each snapshot (MOSES in \cite{greene2010}), Louvain in \cite{louvain}, etc.) and for community matching (Jaccard Coefficient in \cite{greene2010}, Core nodes in \cite{Wang2008}, etc).

In recent years, several authors have proposed methods based on a different approach, allowing to work on dynamic graphs provided as a stream. In this case, there are too many modifications of the network to run a complete algorithm at each step. Therefore, these methods update communities found at previous steps based on local rules. Below, we introduce examples of such methods. More details can be found in \cite{cazabet2014dynamic}.
\begin{itemize}

\item Xie et al.\cite{xie2013b} extended LabelRank \cite{xie2013a} algorithm which is a  stabilized and deterministic variant of Label propagation algorithm \cite{xie2011} to deal with evolving communities in dynamic networks. The extended algorithm called LabelRankT is based on a conditional update rule by which only nodes involved in change between two consecutive snapshots are updated.

\item Nguyen et al.\cite{nguyen2011} proposed AFOCS, an adaptive framework for detecting, updating and tracing the evolution of overlapping communities in dynamic mobile networks. During the initialisation step, AFOCS identifies all possible basic network communities which represent the densely connected part of the network, whose internal density is greater than a certain level, and merge those with the highest overlaps with each other. In a second step, AFOCS adaptively update the community structure, as the dynamic network evolves in time.

\item Cazabet and Amblard \cite{cazabet2011} proposed an online algorithm called iLCD. In this work, the dynamic network is considered as a sequence of events (adding or removing edges). iLCD is using a multi-agent system: each community is an agent on the network, which can integrate or reject nodes. The agents are bounded by a certain number of operating rules, like updating existing communities, creating new communities or merging similar ones. Communities can be updated at each apparition or deletion of links.

\item Rossetti et al.\cite{rossetti2016} defined TILES, which also proceeds in a streaming fashion, i.e., dynamics of the network is described as flows of interactions (also called perturbations) between users where nodes and edges can be created or removed over time. Each perturbation is considered as a fall of domino tile: every time a new interaction appears in the network, TILES updates the community locally and then propagates the changes to the node surroundings to adjust the neighbors' community memberships. 
\end{itemize}

A weakness of these algorithms is the absence of any guarantee that the communities found represent an optimal solution at the global level, because communities at each step are based on communities found in a previous step by applying a set of local rules. More precisely, these methods suffer from the risk of community drift, in which the solution can be dragged away from an originally relevant solution. Another consequence is that communities found by these algorithms at step $t$ depend on the particular sequence of previous graph modifications: the same graph produced by a different graph's history would yield a different partition.

On the contrary, due to the nature of the definition of communities in CPM, we are able in this article to provide an algorithm that handles a flow of changes with local modifications, while guaranteeing that the same state of the graph will always yield the same community structure.

\section{Dynamic Network Model}
\label{dynamicNetworkModel}
Various temporal models have been proposed to deal with dynamic networks. We distinguish three broad approaches: 

\begin{itemize}
\item \textbf{Aggregated graphs} model the dynamic network as a single static network by aggregating all contacts between each pair of nodes in a single edge. This representation does not allow longitudinal analysis, for instance tracking the evolution of communities.

\item \textbf{Series of snapshots} model the evolving network through a series of snapshots, each of which is a static network representing contacts that exist at the corresponding time, or during the corresponding time window. The main issue of this approach is to determine the 'right' number of time windows, i.e., the temporal granularity. Tracking communities across network sequences can be difficult if important temporal information is lost between snapshots.

\item \textbf{Temporal networks} conserve all known temporal information. There are two main models: series of contact and interval graph \cite{holme2012}. In a sequence of contact, interaction is represented as a triple $(i, j, t)$ where $i$ and $j$ are the interacting entities and $t$ is the time when the relationship is activated. In an interval graph, interaction is represented as a quadruplet $(i,j,t,\delta t)$ which means that $i$ is involved in contact with $j$ from $t$ to $\delta t$. In these models, only the temporal information about interactions is represented, there is no temporal information about nodes. 
\end{itemize}

In the following, we introduce our own formalism for evolving graphs, which is better suited to deal with \textit{stream graphs}, i.e., graphs whose modifications occur as a flow, not necessarily known \textit{a priori}. This formalism has the same expressivity as interval graphs.

\subsection{Stream graph}
\label{streamGraph}

Networks are often represented by a graph $G=(V, E)$, where $V$ is the set of nodes and $E$ is the set of edges between nodes. We represent dynamic graphs as an ordered sequence of events, which can be node addition, node removal, edge addition or edge removal. We use the following notations:

\begin{itemize}
\item {\em Inserting or removing a node} is represented as triples $(v,e,t)$,  where $v$ is the node, $e$ is the event observed among $\{+,-\}$(insert ($+$) or remove($-$)), and $t$ is the time when the event occurs.

\item {\em Inserting or removing an edge} is represented as quadruplets $(u,v,e,t)$, where $u$ and $v$ are endpoints of the edge, $e$ is the event observed among $\{+,-\}$( insert ($+$) or remove($-$)), $t$ is the time when the event occurs.

\end{itemize}

Note that this formalism, for edges, is identical in nature to an interval graph, but is more convenient for stream algorithms, as new operations can be added at the end of the ordered sequence of events without affecting previous ones.

\section{OLCPM Framework}
\label{OLCPMsection}

Our framework comprises two main steps. First, we propose to adapt the classical algorithm CPM \cite{palla2005} for static overlapping community detection to deal with evolving networks. We propose an online version of CPM called OCPM (Online CPM). This algorithm is based on analyzing the dynamic behaviors of the network, which may arise from inserting or removing nodes or edges, i.e., every time a change is produced in the network, we update locally the community structure alongside the involved node or edge. 

As stated earlier, CPM may not cover the whole network, i.e., some nodes have no community membership. To deal with this problem, we assume that the communities corresponding to OCMP contain core nodes, and we propose a way to discover the community peripheral nodes. In the second step of our framework, we extend OCMP using label propagation method and we propose OLCPM (Online Label propagation CPM). These proposals will be presented in detail in the next section.
\begin{figure} [h]
\centering
\begin{subfigure}{\linewidth}
    \centering
    \includegraphics{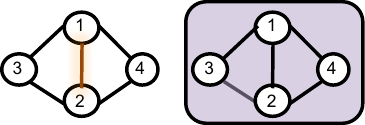} 
    \caption{Example with k=3}
 \end{subfigure}

\begin{subfigure}{\linewidth}
    \centering
    \includegraphics{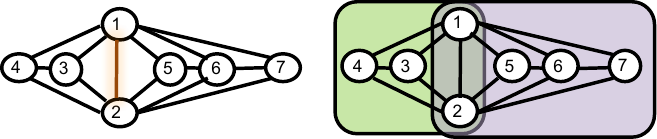} 
    \caption{Example with k=4}
 \end{subfigure}

\caption{Examples of adding an edge with both endpoints outside any community. $(a)$ Example for $k=3$: when the edge$(1, 2)$ is added, a new community $\{$1, 2, 3, 4$\}$ is created from two adjacent $k$-cliques $\{1, 2, 3\}$ and $\{1, 2, 4\}$. $(b)$ Example for $k=4$: the insertion of edge$(1, 2)$ leads to the creation of two communities $\{ 1, 2, 3, 4\}$ and $\{1, 2, 5, 6, 7\}$ from respectively two groups of not-adjacent $k$-cliques $\{\{1, 2, 3, 4\}\}$ and $\{\{1, 2, 5, 6\}$,$\{1, 2, 6, 7\}\}$}.
\label{fig:AddExternalEdge}
\end{figure}

\begin{figure} [h]

\centering
\begin{subfigure}{\linewidth}
    \centering
    \includegraphics{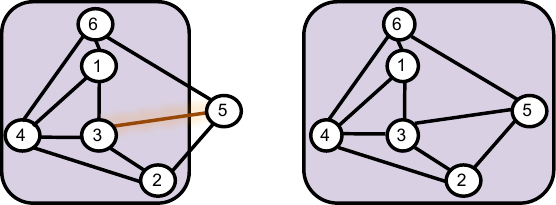} 
    \caption{Simple grow}
 \end{subfigure}

\begin{subfigure}{\linewidth}
    \centering
    \includegraphics{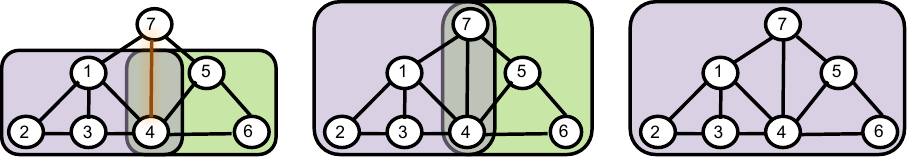}
    \caption{Grow and merge}
 \end{subfigure}

\begin{subfigure}{\linewidth}
    \centering
    \includegraphics{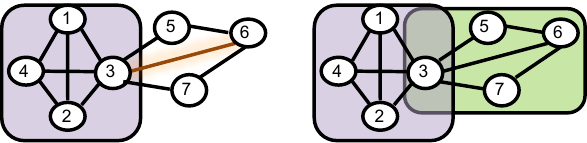} 
    \caption{New community}
 \end{subfigure}
 
\caption{Example of adding an edge with an external endpoint and internal one(for $k=3$). (a) The community $\{1, 2, 3, 4, 6 \}$ grows with node $5$ when adding edge $(3, 5)$. (b) When the edge $(4,7)$ is added, the communities $\{ 1,2,3,4\}$ and $\{4,5,6\}$ grow with node $7$, and then merged. The resulting community takes the identity of the one that contains more nodes.(c) By adding edge $(3,6)$, a new community $\{ 3,5,6,7\}$ is created.}

\label{fig:OneEexternalEndpoint}
\end{figure}

\begin{figure} [h]

\begin{subfigure}{\linewidth}
    \centering
    \includegraphics{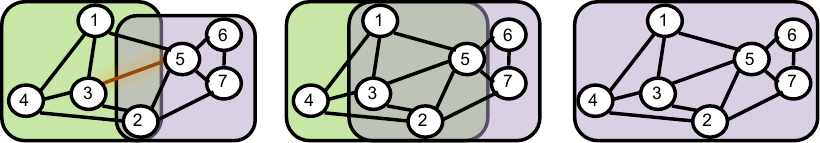}
    \caption{Grow and Merge}
 \end{subfigure}
 
 \begin{subfigure}{\linewidth}
    \centering
    \includegraphics{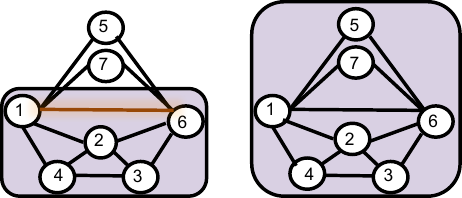} 
    \caption{Grow}
 \end{subfigure}

\caption{Examples of adding an edge with two internal endpoints(k=3). (a) The communities $\{1,2,3,4\}$ and $\{2,5,6,7\}$ grow with the nodes of adjacent $k$-cliques $\{ \{1,3,5\},\{2,3,5\}\}$ formed when adding the edge $(3,5)$, and then merged. (b) The community $\{1,2,3,4, 6\}$ grows with the nodes of adjacent $k$-cliques $\{ \{1,7,8\},\{1,5,8\},\{1,2,8\}\}$ formed when adding the edge $(3,5)$.}
\label{fig:TwoInternalEndpoints}
\end{figure}

\begin{figure} [h]
\centering

\begin{subfigure}{\linewidth}
    \centering
    \includegraphics[width=.3\linewidth]{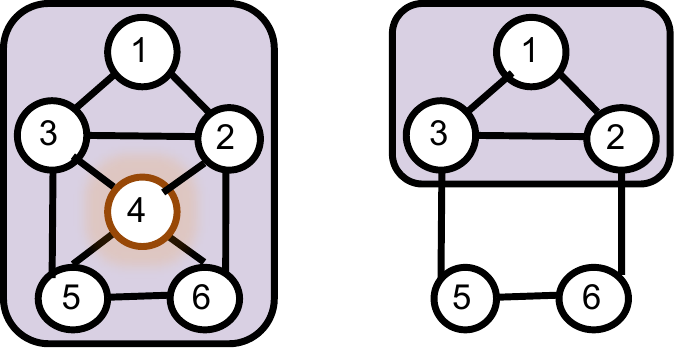}
    \caption{Shrink}
 \end{subfigure}
 
 \begin{subfigure}{\linewidth}
    \centering
    \includegraphics[width=.4\linewidth]{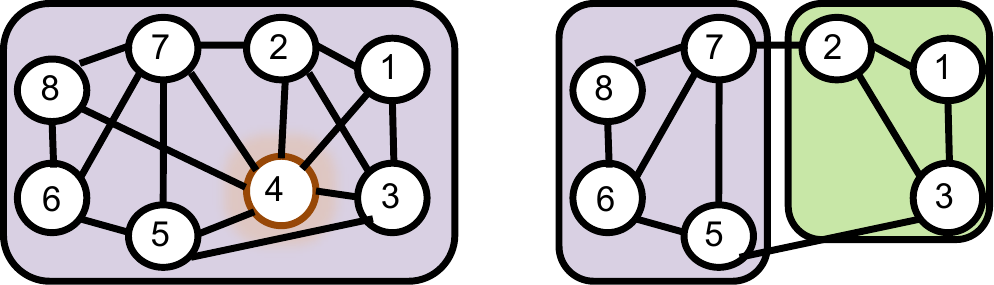} 
    \caption{Shrink and Split}
 \end{subfigure}
 
 \begin{subfigure}{\linewidth}
    \centering
    \includegraphics{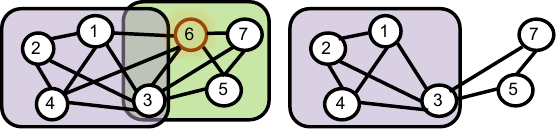} 
    \caption{Death}
 \end{subfigure}
 
\caption{Example of removing internal node (k=3 for (a) and (b), k=4 for (c)). (a) When removing the node $4$, the members $\{4,5,6\}$ leaves out the community $\{1,2,3,4,5,6\}$.(b) When removing the node $4$, the community $\{1,2,3,4,5,6,7,8\}$ shrinks, i.e., it loses this node and all its edges, and then splits into two communities: $\{5,6,7,8\}$ and $\{1,2,3\}$. (c)By removing the node $6$, the community $\{1,2,3,4\}$ shrinks and the community $\{3,5,6,7\}$ dies}
\label{fig:DeleteInternalNode}
\end{figure}

\begin{figure} [h]

 \begin{subfigure}{\linewidth}
    \centering
    \includegraphics{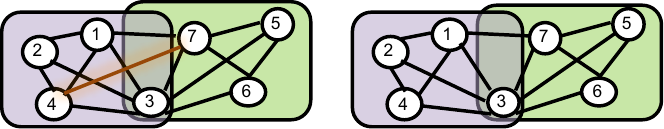}
    \caption{No change in the community structure}
 \end{subfigure}
 
  \begin{subfigure}{\linewidth}
    \centering
    \includegraphics{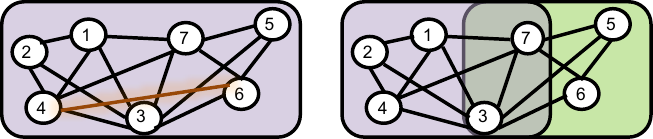}
    \caption{Community split}
 \end{subfigure}
 
\caption{Examples of removing internal edge (k=4). (a) The community structure doesn't change when removing the edge $(4, 7)$. (b) When removing the edge $(4, 6)$, the community splits into two small communities, each of which contains a group of adjacent $k$-cliques in the original community.}
\label{fig:RemoveInternalEdge}
\end{figure}


\subsection{OCPM: Online Clique Percolation Method}

This section proposes the first step of our framework OLCPM, an online  Clique Percolation Method (OCPM). This method takes two inputs: 
\begin{itemize}
\item $SE$, chronologically ordered sequence of events which models networks modification, following the format: $(n, e, t)$ or $(i, j, e, t)$ as defined in section \ref{streamGraph}
\item the parameter $K$, which determines the clique size; it is an integer value greater than or equal to 3 
\end{itemize}

The OCPM method  maintains after each modification three elements: 
\begin{itemize}
	\item $G(V, E)$ the current state of the network
	\item $AC$ the set of currently alive communities 
	\item $DC$ the set of dead communities
\end{itemize}

It is therefore possible to know the community structure status at every network modification step. 

\subsubsection{Definition of the OCPM algorithm}
Note: To facilitate the readability of the paper, we decided to put all formal algorithms in the \textbf{Appendix}, and to only include the rationale of these algorithms in the body of the article. Please refer to the \textbf{Appendix} for further details.

\bigbreak

The core of the OCPM algorithm can be defined by an algorithm that updates the current state of all variables according to a sequence of events $SE$, as detailed in Algorithm \ref{algo:OCPM}. The task carried out by the algorithm depends on the type of event encountered:

\begin{itemize}
\item \textbf{Add a new node}: adding an isolated node $n$ has no influence on the community partition. In this case, only $n$ is added to the graph $G$ and no other action is performed until the next event. 
\item \textbf{Add a new edge}: when a  new edge $(i,j)$ appears, we add this edge to the graph $G$. 
According to the type of edge, we distinguish two cases:

	\begin{itemize}

	\item When inserting an external edge, i.e., both its endpoints are outside any community, we check if one or more new $k$-cliques (KCliques() function Algorithm \ref{KCliques}) are created. If it is the case, we gather all adjacent $k$-cliques one to the other. Then, for each group of adjacent $k$-cliques, we create a single community. Figure \ref{fig:AddExternalEdge} shows two examples of adding external edges and the changes it brings to the community structure. (See Algorithm \ref{Algo:AddExternalEdge})
	\item In all other cases, i.e., when a new edge appears with one or two internal extremities, we check all $k$-cliques created when adding this edge and not belonging to any community. Then, all adjacent $k$-cliques are grouped together and for each group, we check if there are other adjacent $k$-cliques included in any community to which belongs any node in this group. If they exist, the corresponding communities will grow with the nodes of this group and they can eventually be merged (Merge()function Algorithm \ref{Merge}). Otherwise, a new community appears containing nodes of this group. Figures \ref{fig:OneEexternalEndpoint} and \ref{fig:TwoInternalEndpoints} depict some examples of adding edges with one or two internal endpoints and the changes to the community structure. (Algorithm \ref{Add_internal_edge})
	\end{itemize}

\item \textbf{Delete node}: In this case, we remove the node from the graph G, and all its edges are removed as well. If the node is external, i.e., it doesn't belong to any community, the community structure is not affected and no action is performed until the next event. When the removed node belongs to one or more communities, we check for each community to which this node belongs whether it still contains at least a $k$-clique after the node is removed. This community dies if it loses all $k$-cliques(see figure (c) \ref{fig:DeleteInternalNode}). Otherwise, the community shrinks, i.e., it loses this node and all its associated edges. Here, we distinguish two cases:
\begin{itemize}
\item The community may remain coherent and the community structure doesn't change(see figure (a) \ref{fig:DeleteInternalNode} ).
\item The community may become disconnected and therefore, it will be break up into small communities (see figure (b) \ref{fig:DeleteInternalNode}).
\end{itemize}
The split function (Algorithm \ref{Split}) deals with these two cases. After the community shrinking, its structure is recalculated keeping the principle of CPM -checking all maximal cliques of size not less than $k$. The resulting community having the largest number of nodes keeps the identity of the original one, where the others have new identities.

The Algorithm \ref{Remove_internal_node} describes this case.

\item \textbf{Delete edge}: First, we remove the edge from the graph G. The removal of an edge with two endpoints belonging to the same community(ies)(called internal edge) follows the same mechanism as internal node removal: the communities to which belong the two extremities of this edge may split or die. For each of them, we check whether it still contains $k$-cliques. If so, we use the function Split (Algorithm \ref{Split}) to check whether or not the community is divided into smaller parts. Otherwise, this community dies (see Algorithm \ref{Remove_internal_edge}). Figure \ref{fig:RemoveInternalEdge} shows two examples of removing internal Edge and the changes that it brings to the community structure.

For all other types of edges, the community structure doesn't change. 

\end{itemize}

Here, we detail some functions used in our algorithm:

\begin{itemize}

\item \textbf{Kcliques}(): (Algorithm \ref{KCliques}) This function takes a set of nodes SN as input parameter and returns all maximal cliques of size not less than $k$ containing this set. In order to optimize the performance of our algorithm, $k$-cliques are locally launched in the subgraph including the set $SN$ and all common neighbors among its members. 

\item \textbf{Merge}(): (Algorithm \ref{Merge}) This function is used for merging adjacent communities. The resulting community takes the identity of the one with the highest number of nodes. 

\item \textbf{Split}(): (Algorithm \ref{Split}) This function is used for splitting a community if possible. It takes as input a community and creates from it one or more communities.  We proceed as follows: first, we identify all maximal cliques of size not less than $k$ in this community and we aggregate adjacent $k$-cliques with each other. Then, for each of the aggregated $k$-cliques, we create a new community. The community which has the largest number of nodes keeps the identity of the original one.
\end{itemize}

Table \ref{tab:Actions} summarizes the actions which can be carried out by OCPM according to graph events.

\begin{table} [!h] 
\centering
\begin{tabular}{|l|l|l|}
\hline
\multicolumn{2}{|l|}{\textbf{Event}} & \textbf{Actions} \\
\hline
\multicolumn{2}{|l|}{Add new node} & - \\
\hline
\multirow{2}{*}{Add new edge} & External &  Birth \\ \cline{2-3}
                  & Other & Grow+[Merge], Birth \\ 
\hline
\multirow{2}{*}{Delete Node} & External & - \\ \cline{2-3}
                  & Internal & Shrink+[Split], Death \\
\hline
\multirow{2}{*}{Delete Edge} & Internal & Split, Death \\ \cline{2-3}
                  & Other & - \\
\hline
\end{tabular}
\caption{Actions that can be performed according to graph events. Brackets denotes events that can only follow the preceding community event.}\label{tab:Actions}

\end{table}

\subsubsection{Complexity of the algorithm}

Instead of computing all $k$-cliques for the whole network at each event occurring in the network, OCPM updates the community structure on the local scale, and thus only the community structure alongside the node or the edge involved in the event is recomputed. For certain events, like adding or deleting an isolated node or deleting an external edge, the community structure doesn't change and hence, the computational time saving reaches its maximum. For instance, if we have $n$ $k$-cliques when such event is produced, the computational time savings will be $n$ times the average time for calculating $k$-cliques. For other events, the computational time saving is also significant. See section \ref{empComplexity} for an empirical evaluation of the complexity.

\subsubsection{Community tracking process}
One of the difficulties when tracking the evolution of communities is to decide which community is a continuation of which. Our framework allows a trivial matching in the case of \textit{continuation} (no merge or split) of communities. In the case of merge and split, deciding which community keeps the original identity is a well-known problem with no consensus in the literature \cite{cazabet2014dynamic}. In OCPM, we took the simple yet reasonable decision to consider that the \textit{largest} community involved in a merge or split have the same identifier as the merged/split one. This strategy can be replaced without altering the algorithm logic. A more advanced process could be added to solve problems of \textit{instability}, e.g. communities merging and quickly splitting back to their original state.

\subsection{OLCPM: Online Label propagation CPM}

This section describes the second step of our framework. A post-processing based on label propagation is set out on the output communities of OCPM to discover the peripheral nodes. This module is called OLCPM (Online Label propagation CPM).  

There is a twofold reason for using a post-process extending core-communities found by OCPM:
\begin{itemize}
	\item In a network evolving at fast path, one can update core-communities efficiently after each event, and run the post-process only when the current state of communities needs to be known, thus saving computation time
	\item It is known that the periphery of communities is often not well defined and unstable. As seen earlier, and because OCPM is deterministic and it searches for core-communityies, it reduces this instability problem. By using the label propagation mechanism only as a post-process for analysis, communities at $t$ do not depend on the periphery of communities that might have been computed at $t-1$, but only on the stable part found by OCPM.
\end{itemize}

\subsubsection{OLCPM algorithm}

First, each core-community (community found by OCPM) spreads to neighboring peripheral nodes (nodes not covered by OCPM) a label containing its identity and a weight representing the geodesic distance (the length of the shortest path) between this neighboring node and any other node in the core-community. Each peripheral node has a local memory allowing the storage of many labels. Label propagation process is based on breadth-first search (BFS). When all labels have been shared, nodes are associated with all communities with which they have the shortest geodesic distance. Note that nodes can, therefore, belong to several communities, if they are at the same distance of community found by OCPM. This algorithm is defined formally in Algorithm \ref{algo:OLCPM}.

\begin{figure*}[h!]
    \centering
     \begin{subfigure}{\linewidth}
     \centering
    \includegraphics{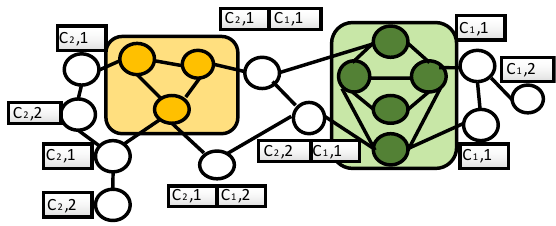}
    \caption{Label spreading step}
 \end{subfigure}
  \begin{subfigure}{\linewidth}
  \centering
    \includegraphics{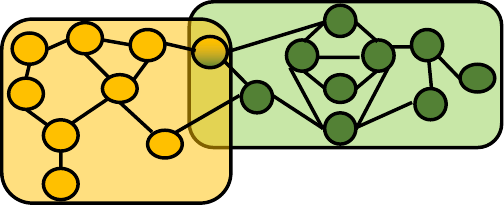}
    \caption{Community structure after label analysis (k=3)}
 \end{subfigure}
\caption{Peripheral community updates by OLCPM. (a) Label spreading step. (b) Community structure after label analyses (for K=3). Green nodes are members of the community $C1$; Yellow nodes are  members of the community $C2$; uncolored nodes have no affiliation.}
\label{fig:OLCPM}
\end{figure*}

Figure \ref{fig:OLCPM} presents an illustration of this process.

\section{Experiments}
\label{Experiments}
In this section, we begin by evaluating the effectiveness of OCPM algorithm. Thus, we compare the time complexity of OCPM with the dynamic version of CPM \cite{palla2007}. Second, we are interested in the quality of the communities that OLCPM is able to find, considering both synthetic and real-world networks.

\subsection{Measuring OCPM complexity gain for highly dynamic networks}
\label{empComplexity}
In this section, we compare the empirical complexity of the original dynamic version of CPM (hereafter, DyCPM)\cite{palla2007} and our proposed version (OCPM). We generate synthetic dynamic networks, and compare how the running time of both algorithms vary with the properties of the network and of its dynamic. Note that we compare OCPM only with CPM because both algorithms try to solve the \textit{same problem}, i.e, they have the same definition of communities. Other streaming algorithms introduced in section \ref{relatedWork} have an \textit{ad hoc} definition of communities introduced together with the method, and does not have the same properties, such as being deterministic and not being dependent on the network history. Their complexity is, in theory, similar to the one of OCPM (local updates at each modification).

\subsubsection{Generation of dynamic networks with community structure}
We propose a simple process to generate dynamic networks with realistic community structure. First, a static network is generated using the LFR benchmark \cite{lancichinetti2009a}, the most used benchmark for community detection. Then, for this network, we generate a step by step evolution. In order to conserve the network properties (community structure, size, density), we define an \textit{atomic modification} as the following process:

\begin{enumerate}
\item Choose randomly a planted community as provided by LFR
\item Select an existing edge in this community
\item Select a pair of nodes without edges in this community
\item Replace the selected existing edge by the selected not-existing one.
\end{enumerate}

We define a step of evolution as the combination of $a$ atomic modifications. In order to test the influence of the number of modifications between steps, we test different values of $a$.

Note that we use synthetic networks instead of real networks at this step since: 
\begin{itemize}
	\item We are only interested in measuring time complexity of algorithms. Synthetic networks are mostly criticized for having unrealistic community structures, while here we are mainly interested in the size and rate of evolution of the networks.
	\item It allows controlled experiments. With real evolving networks, changes in the structure/size of the network could affect computation time at each step, and we could not control the number of modifications between snapshots, or vary the size of networks while keeping constant properties.
\end{itemize}

\subsubsection{Experimental process}

The LFR benchmark \cite{lancichinetti2009a} is, as of today, the most widely used benchmark to evaluate community detection methods. It is known to generate realistic networks with heterogeneous degrees and community sizes.

It has the following parameters : $N$ is the network size, $k$ is the average degree of nodes, $kmax$ the maximum degree, $t1$ and $t2$ are power-law distribution coefficients for the degree of nodes and the size of community respectively, $\mu$ is the mixing parameter which represents the ratio between the external degree of the node with respect to its community and the total degree of the node, $minc$ and $maxc$ are the minimum and maximum community size respectively, $On$ is the number of overlapping nodes , $Om$ is the number of community memberships of each overlapping node. 

In order to obtain realistic networks, we first generate an original network with $n$ nodes using the LFR benchmark, with fix parameters $k=7$, $maxk=15$, and $\mu=0.4$. Other parameters stay at their default values. In order to test the influence of the network size, we test different values of $n$.

\begin{figure} [h!]
\begin{center}
\includegraphics[width=0.8\linewidth]{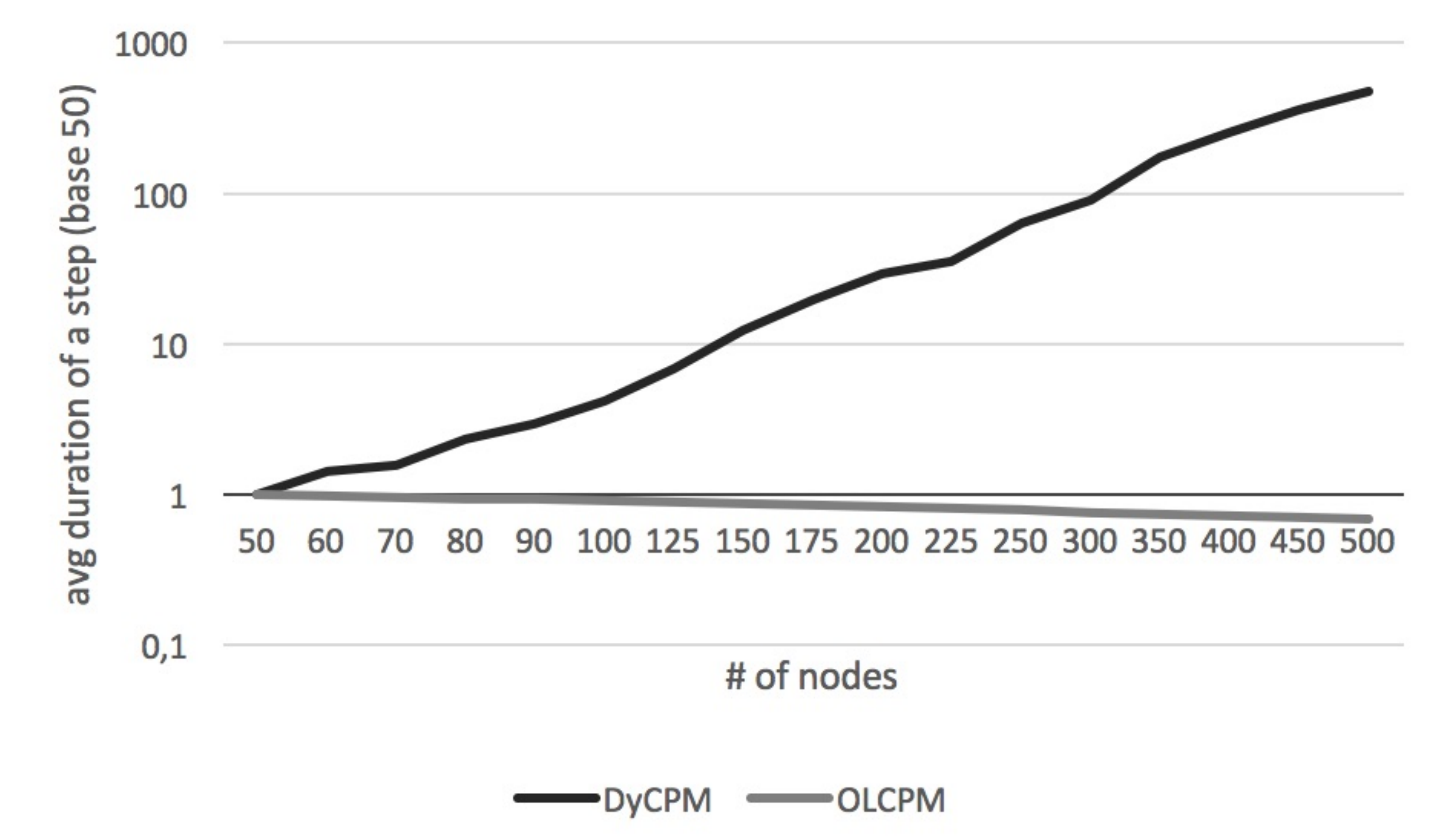}
\end{center}
\caption{Evolution of time complexity when varying the size of the network (number of nodes), and keeping other parameters constant (average node degree, community, size, etc.). DyCPM complexity increases exponentially with the size of the network, while OLCPM one stays constant or slightly decreases. Expressed in base 50, i.e, 10 on the vertical axis means 10 times slower than with 50 nodes.}
\label{fig:time1}
\end{figure}

\begin{figure}[!ht]
\begin{center}
\includegraphics[width=0.8\linewidth]{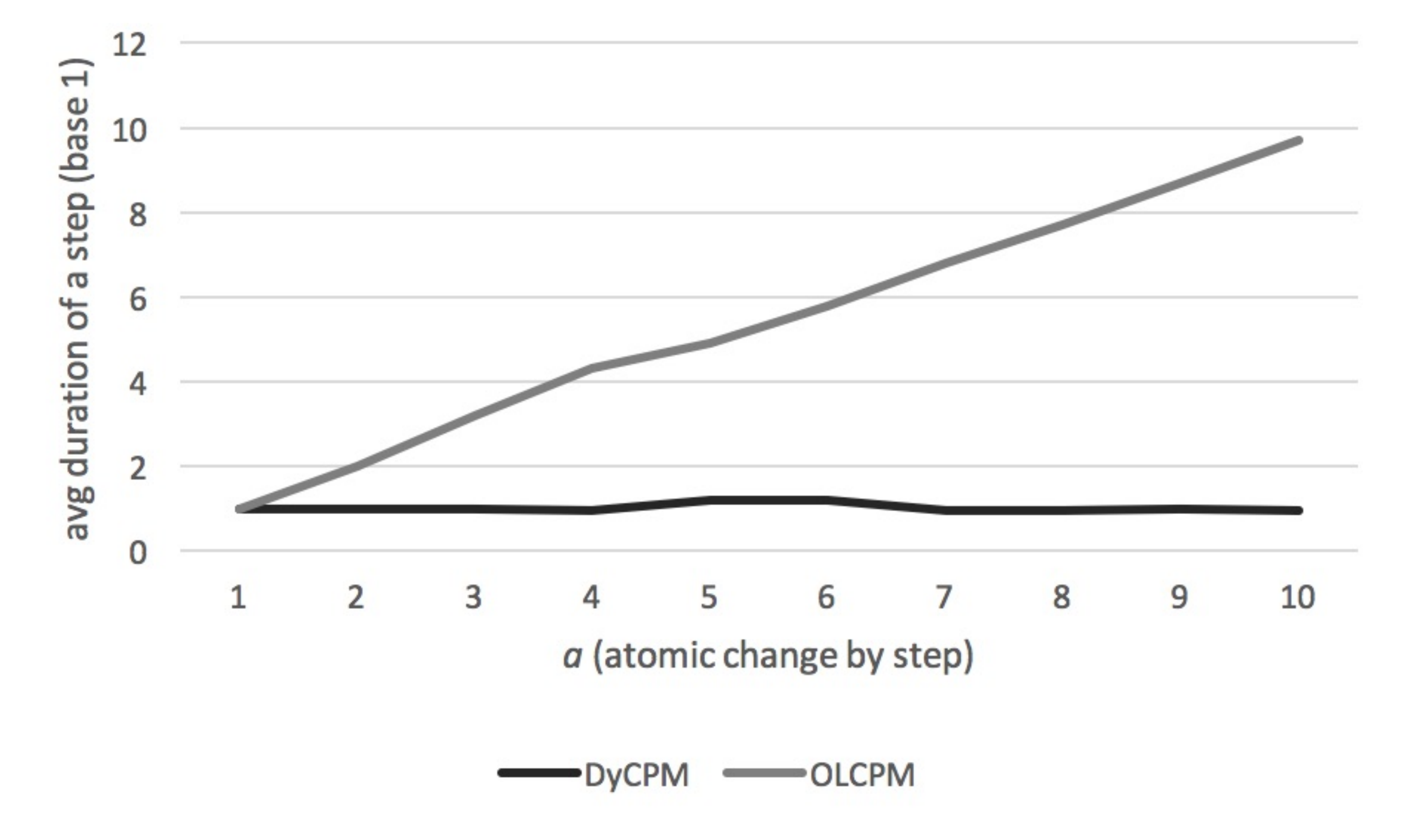}
\end{center}
\caption{Evolution of time complexity when varying the number of atomic changes by step. DyCPM complexity is independent relatively to $a$ while OLCPM's complexity increases linearly with $a$ Time.}
\label{fig:time2}
\end{figure}

As can be seen in figures \ref{fig:time1} and \ref{fig:time2}, the complexity of both algorithms depends on very different parameters. With OLCPM, the time needed to update communities after a modification step does not increase proportionally to the size of the network at any given time, but increases linearly with the number of atomic modifications. 

On the contrary, the complexity of DyCPM depends on the properties of the static network, but not on the number of atomic modifications between steps.

As expected, OLCPM is appropriate to deal with stream graphs, in which modifications are known at a fine granularity, as the cost of each update is low. On the contrary, DyCPM is appropriate to deal with network snapshots, i.e., a dynamic network composed of a few observations collected at regular intervals.

\subsection{Measuring OLCPM communities quality}

To quantify the quality of communities detected by OLCPM framework, we used both synthetic and real-world networks with ground truth community structure. We remind the reader that communities found by DyCPM and OCPM are identical, the difference lies only in the label propagation post-process of OLCPM.

Normalized Mutual Information (NMI) is used as the measurement criterion. This measure is borrowed from information theory \cite{danon2005} and widely adopted for evaluating community detection algorithms. It measures the similarity between a ground truth partition and the one delivered by an algorithm. As the original definition is only well defined for \textit{partitions} (each node belong to one and only one community), a variant of the NMI adapted for \textit{covers} (nodes can belong to zero, one or more communities) have been introduced in \cite{lancichinetti2009b}. This variant is the most used in the literature for comparing overlapping communities. We used the original implementation by the authors \footnote{\url{https://sites.google.com/site/andrealancichinetti/software}}. The NMI value is defined between 0 and 1, with a higher value meaning higher similarity.

\subsubsection{Static Synthetic networks}

We use the LFR benchmark \cite{lancichinetti2009a} to generate realistic artificial networks.  

We use two different network sizes, \textit{small networks}(1000 nodes) and \textit{large networks}(5000 nodes), and for a given size we use two ranges for community size: \textit{small communities}, having between $10$ and $50$ nodes and \textit{large communities}, having between $20$ and $100$ nodes. We generate eight groups of LFR networks. 

In the first four networks, $\mu$ ranges from $0$ to $0.5$ (steps of $0.1$) while $Om$ is set to $100$ for small networks and $500$ for large networks ($5000$ nodes). In the other networks, $\mu$ is fixed to $0.1$ and $On$ ranges from $0$ to $500$ (steps of $100$) for small networks  and from $0$ to $2000$ (steps of $500$) for large networks. All these networks share the common parameters: $k = 10$, $maxk = 30$, $t1 = 2$, $t2 = 1$, $On = 2$. The parameter settings are shown in table \ref{tab:LFRParm}. 

\begin{table} [!h] 
\centering
\begin{tabular}{|c|c|c|c|c|c|c|}

        \hline
        \textbf{Network group ID} & \textbf{N} & \textbf{minc} &	\textbf{maxc} & \textbf{$\mu$} & \textbf{On}\\ 
        \hline
       N1 & 1000 & 10 & 50 & 0-0.5 & 100 \\ 
       \hline
       N2 & 1000 & 20 & 100 & 0-0.5 & 100 \\ 
       \hline
       N3 & 5000 & 10 & 50 & 0-0.5 & 500 \\ 
       \hline
       N4 & 5000 & 20 & 100 & 0-0.5 & 500 \\ 
       \hline
       N5 & 1000 & 10 & 50 & 0.1 & 0-500 \\ 
       \hline
       N6 & 1000 & 20 & 100 & 0.1 & 0-500 \\ 
       \hline
       N7 & 5000 & 10 & 50 & 0.1 & 0-2000 \\ 
       \hline
       N8 & 5000 & 20 & 100 & 0.1 & 0-2000 \\ 
       \hline

\end{tabular}
\caption{LFR parameter setting}\label{tab:LFRParm}

\end{table}

\begin{figure*}[!h]
\centering     
\begin{subfigure}{0.4\linewidth}
    \includegraphics[width=\linewidth]{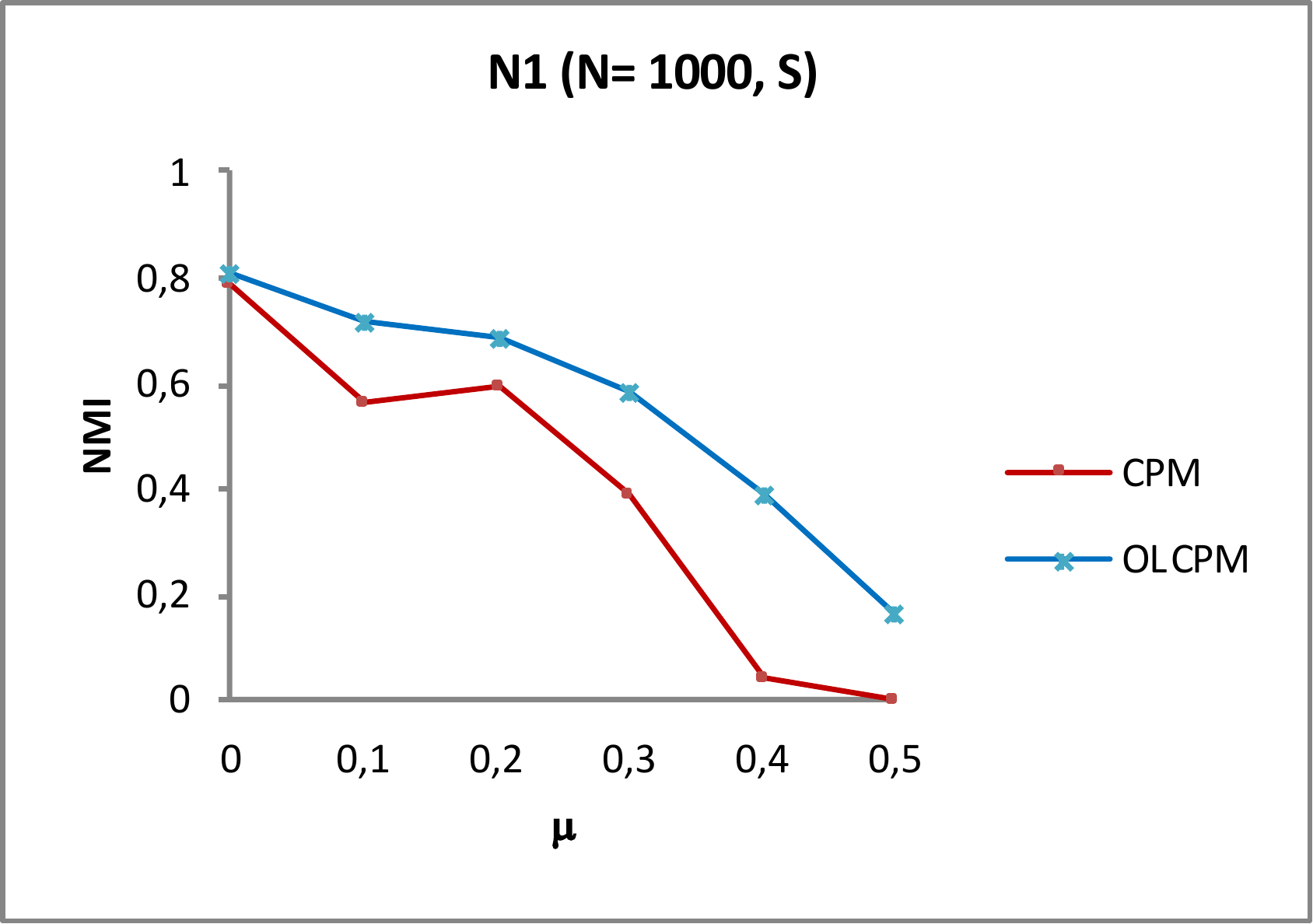}
 \end{subfigure}
     \begin{subfigure}{0.4\linewidth}
    \includegraphics[width=\linewidth]{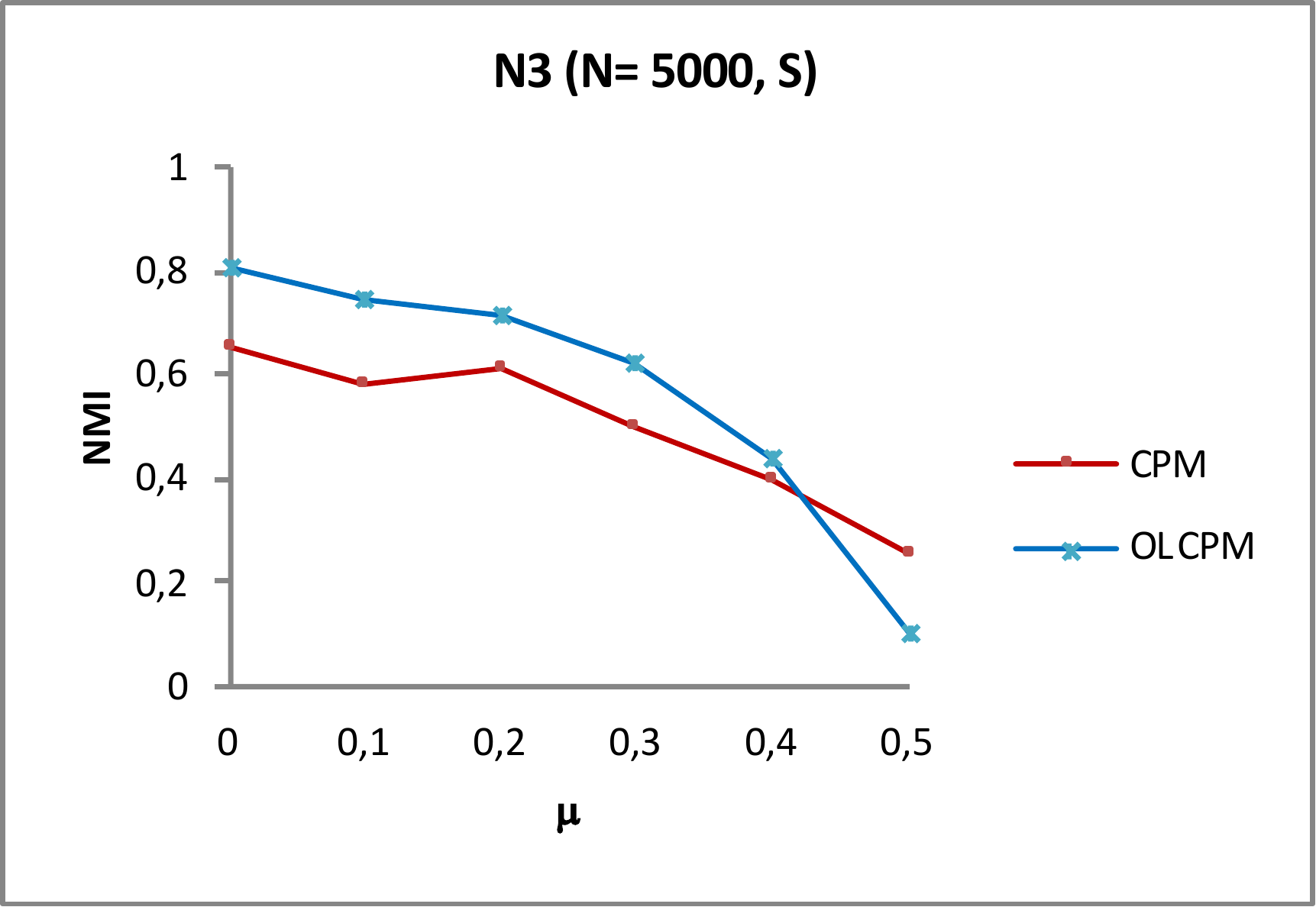}
 \end{subfigure}
 
     \begin{subfigure}{0.4\linewidth}
    \includegraphics[width=\linewidth]{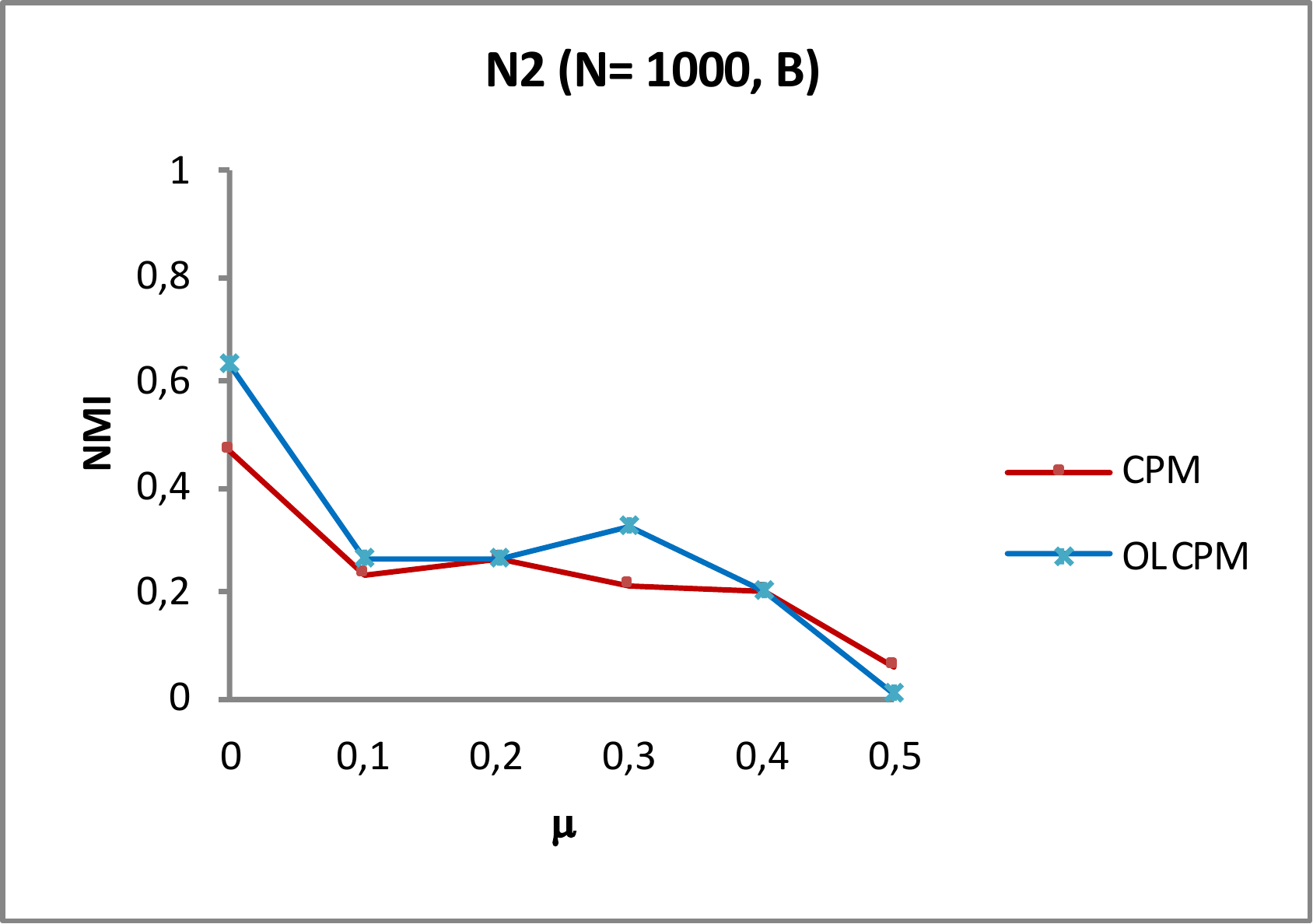}
 \end{subfigure}
     \begin{subfigure}{0.4\linewidth}
    \includegraphics[width=\linewidth]{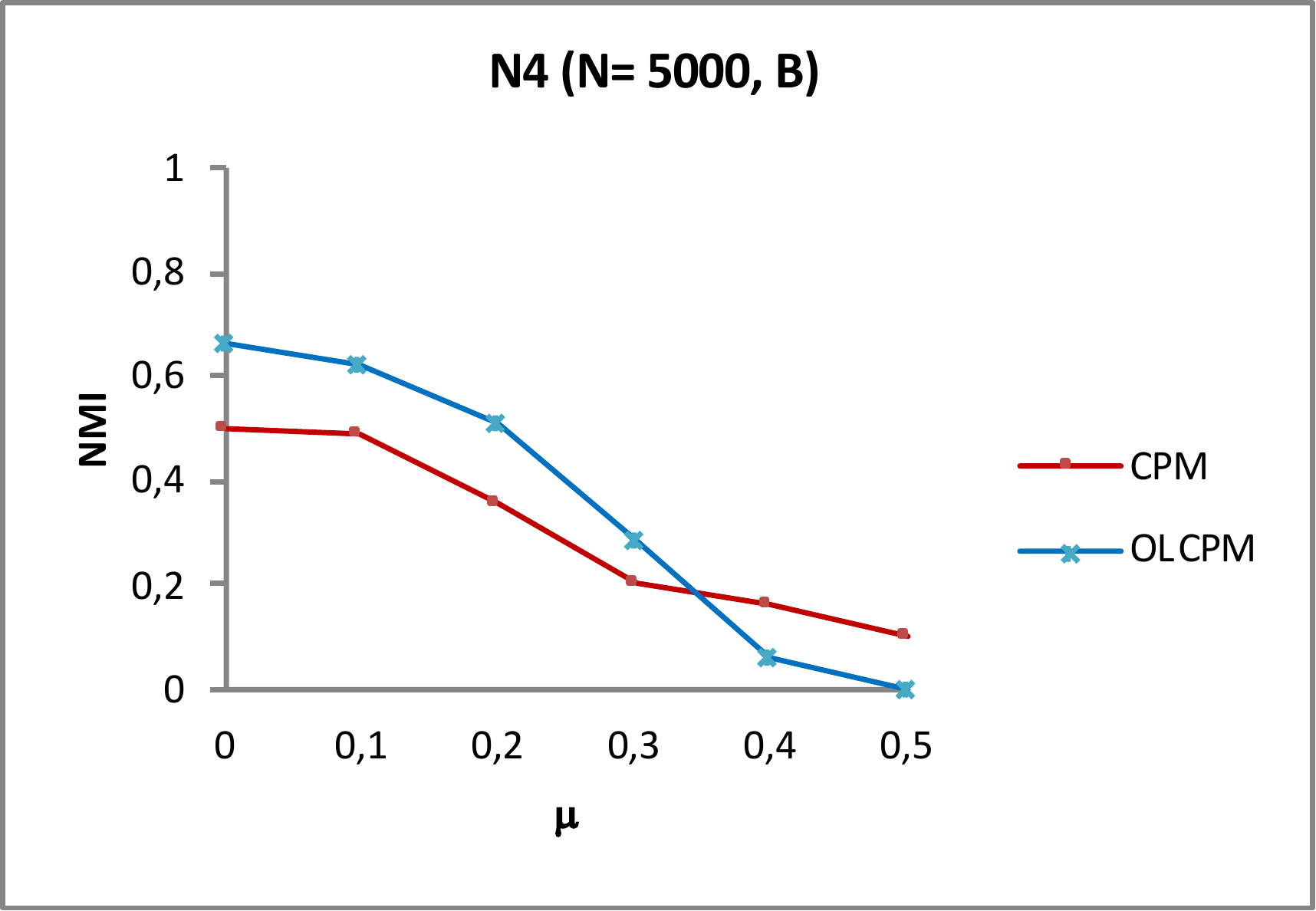}
 \end{subfigure}
 
     \begin{subfigure}{0.4\linewidth}
    \includegraphics[width=\linewidth]{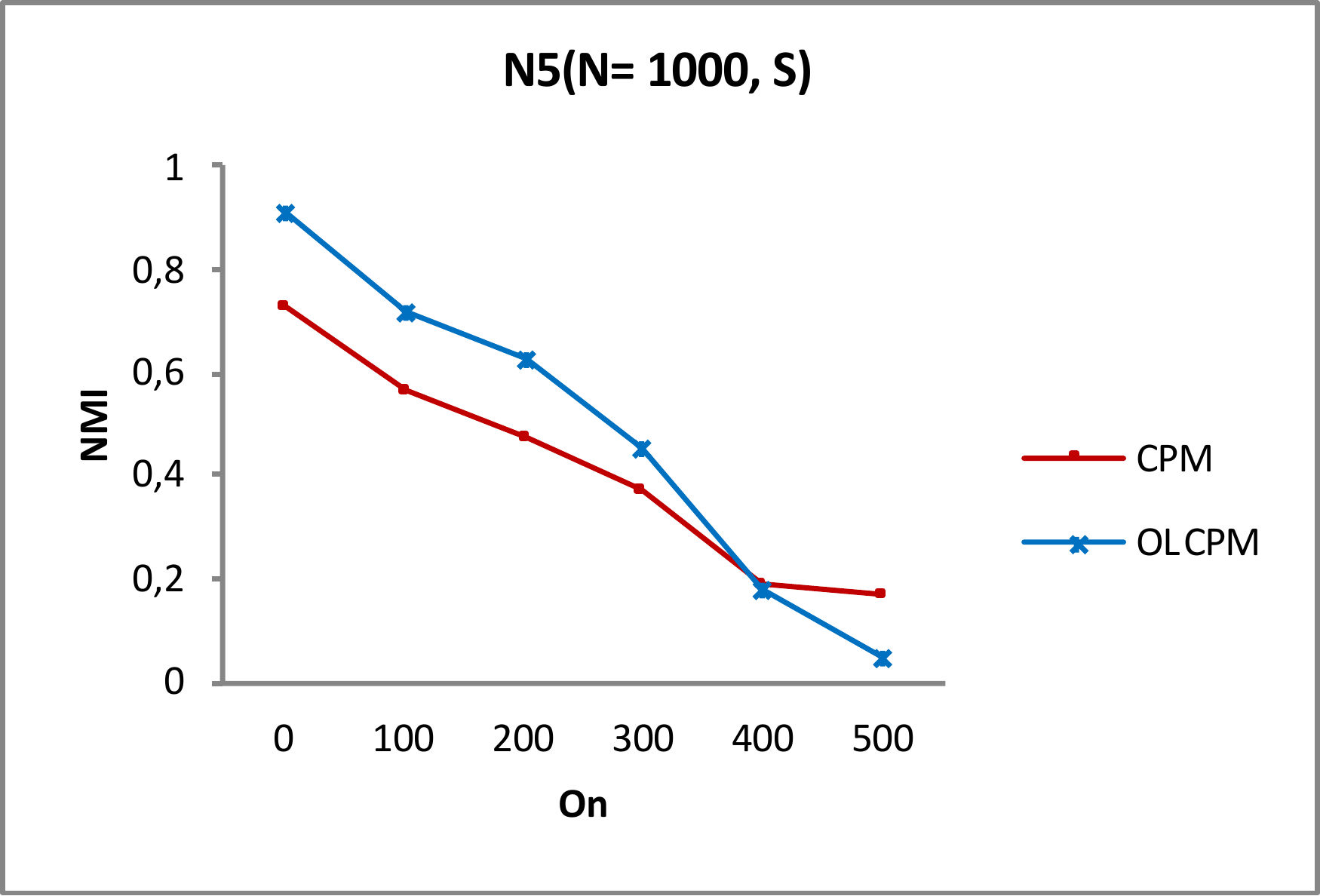}
 \end{subfigure}
     \begin{subfigure}{0.4\linewidth}
    \includegraphics[width=\linewidth]{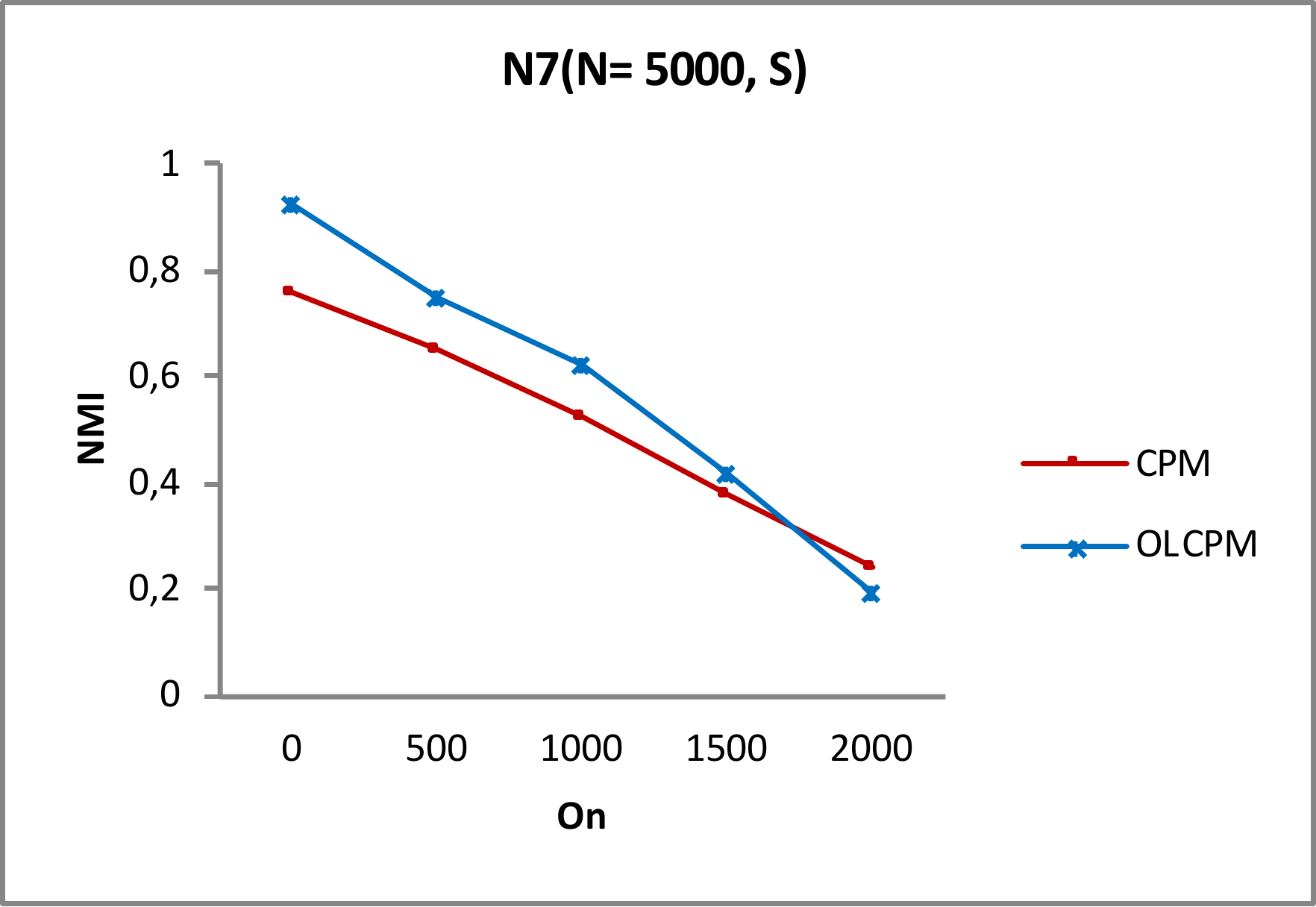}
 \end{subfigure}
     \begin{subfigure}{0.4\linewidth}
    \includegraphics[width=\linewidth]{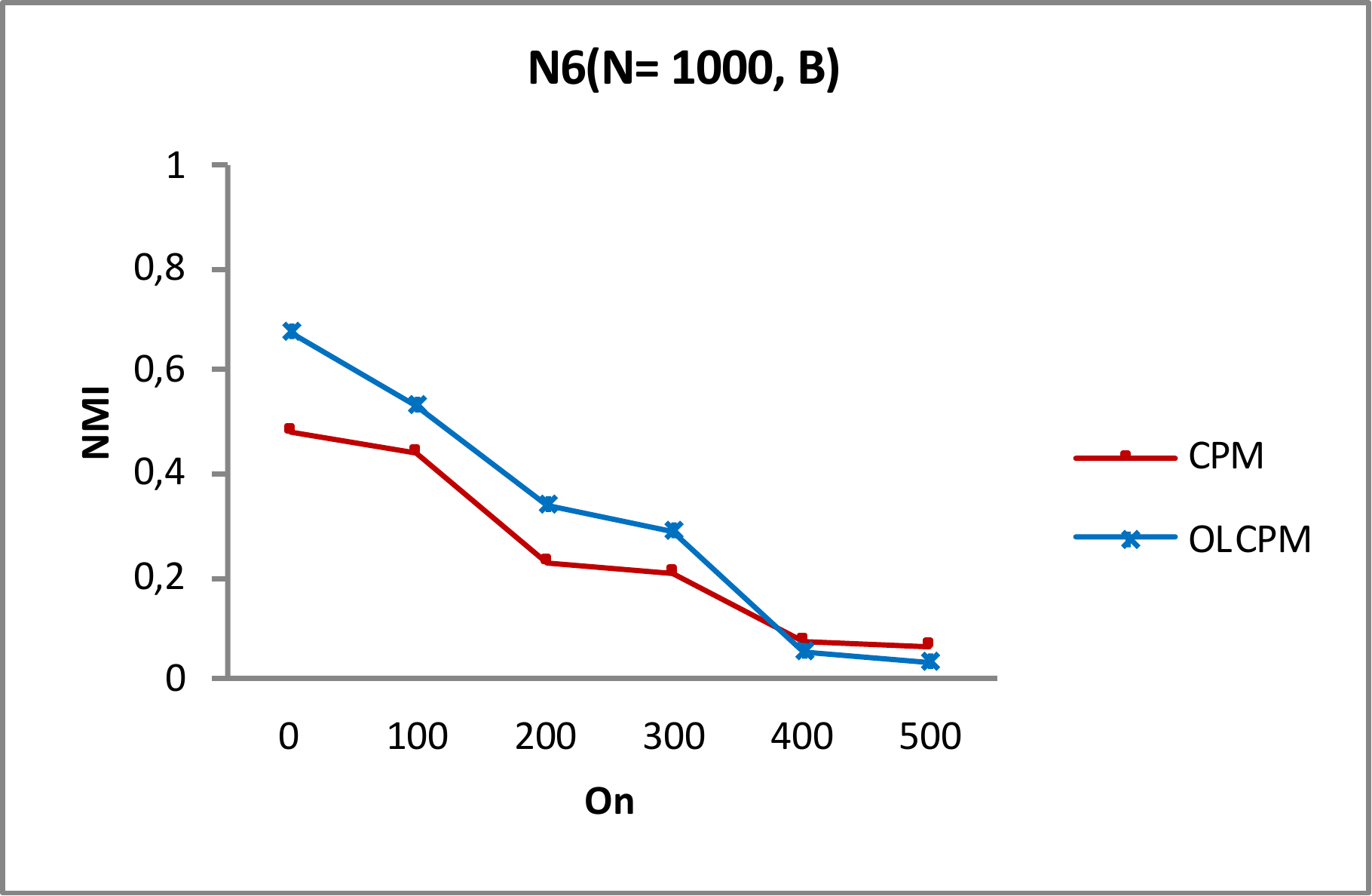}
 \end{subfigure}
     \begin{subfigure}{0.4\linewidth}
    \includegraphics[width=\linewidth]{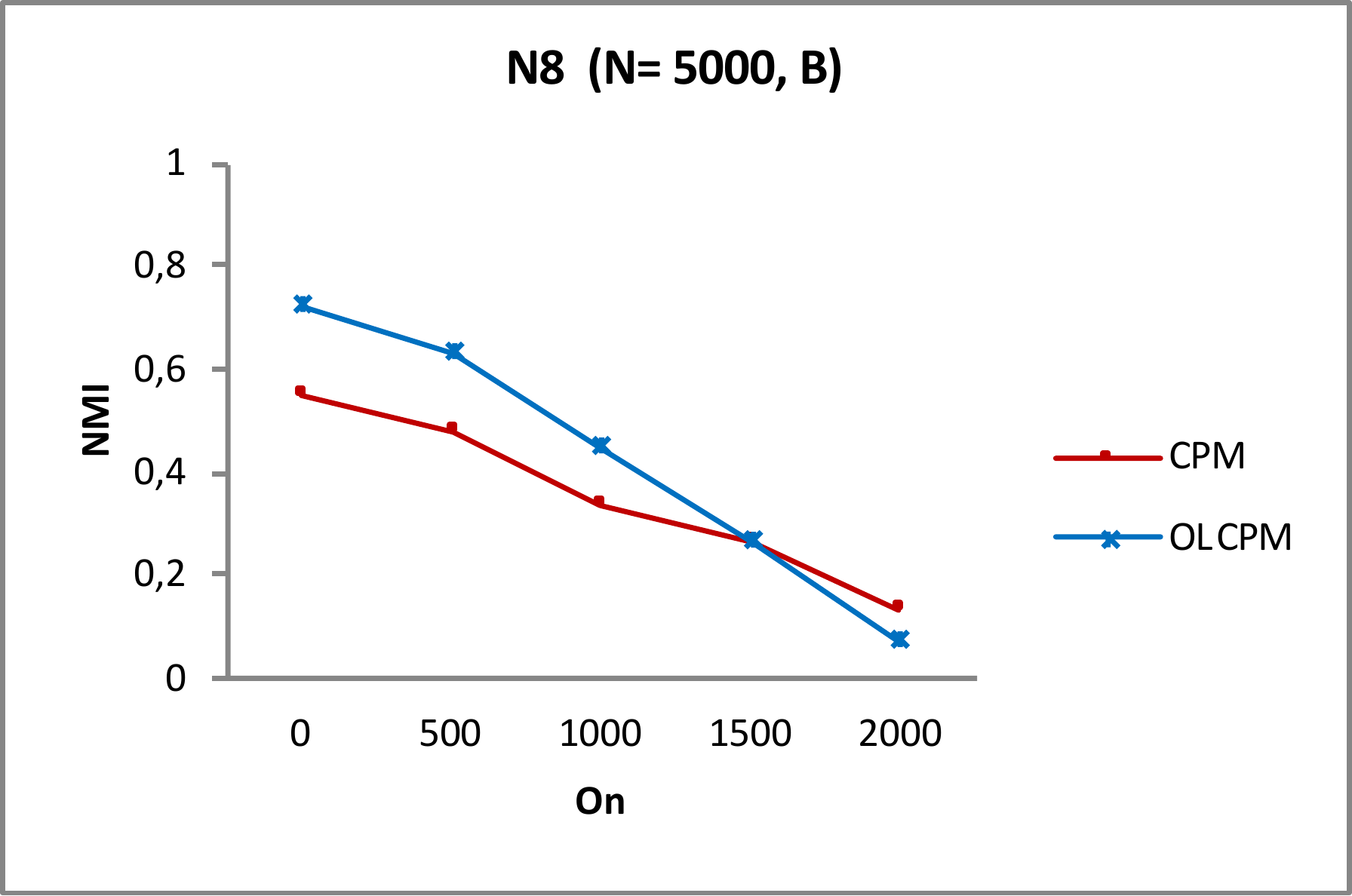}
 \end{subfigure}
\caption{Performance of CPM and OLPM for $k=4$ on the LFR benchmark networks. The plots show the NMI scores as a function  of the mixing parameter $\mu$ (upper half plots) and of the number of overlapping nodes $On$ (lower half plots) for different network sizes (small networks in the left hand plots and large networks in the right hand plots) and different community sizes ($(S)$ ranges from $10$ to $50$ and $(B)$ ranges from $20$ to $100$).}
\label{fig:LFRRes}
\end{figure*}

CPM and OLCPM are run for $k=4$. The NMI values of communities detected by CPM and OLCPM are depicted in figure \ref{fig:LFRRes}. Note that communities found by CPM and OCPM are identical, therefore the observed differences are only due to the post process.

In most cases, OLCPM achieves the highest results, except for the two cases where: (1)  the community structure becomes very fuzzy ( $On >= 400$ for small networks or $On >=1500$ for large networks) or (2) the value of $\mu$ is large (greater than $0.3$). In these cases, OLCPM performs similar or slightly worse than CPM. When the community structure becomes too fuzzy for CPM, the irrelevant core-communities  provided are probably worsened by the post-process. 

As a conclusion, we can consider that in situations in which CPM finds meaningful communities in a network, the proposed post-process improves the solution. 

\subsubsection{Dynamic Real-world networks}

In order to evaluate the community detection results of our framework OLCPM on real temporal networks, we leverage a high-resolution time-varying network describing contact patterns among high school students in Marseilles, France \cite{fournet2014}. The dataset was collected by the SocioPatterns collaboration using wearable sensors, able to capture proximity between individuals wearing them. The dataset was gathered during nine days (Monday to Tuesday) in November 2012. Data collection involved 180 students from five classes. Proximity relations are detected over 20-second intervals. Data collection involved students' classes corresponding to different specializations: 'MP' classes focus more on mathematics and physics, 'PC' classes on physics and chemistry, and 'PSI' classes on engineering studies.  These classes represent the expected ground truth community structure.

We construct a dynamic network composed of 216 snapshots, each corresponding to 1 hour of data. Nodes correspond to students, and there is an edge between two nodes in a snapshot if the corresponding students have been observed in interaction at least once during the corresponding period. (Please refer to the original article \cite{fournet2014} for details about the meaning of \textit{interaction}. To sum up, two students are in interaction if they stand face-to-face at a distance between 1 and 1.5 meters.)

\begin{figure*}[!h]
\begin{center}
\includegraphics[width=\linewidth]{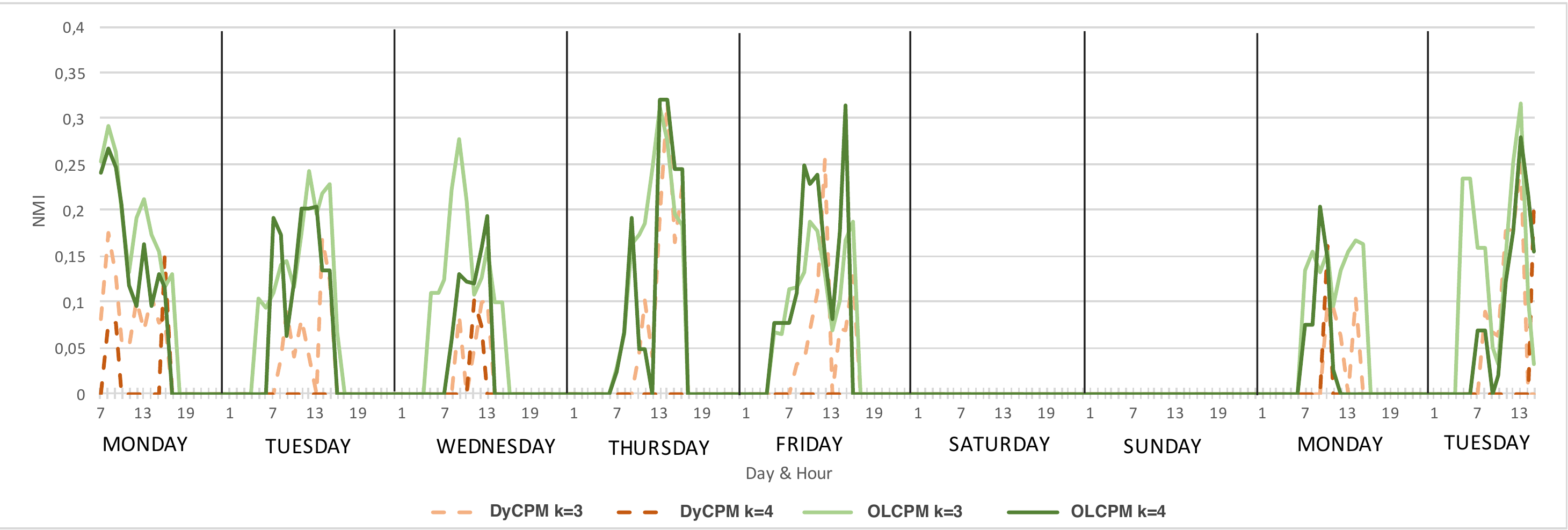}
\end{center}
\caption{NMI values of OLCPM and CPM \cite{palla2005} for $k=3$ and $k=4$ in on SocioPatterns collaboration networks \cite{fournet2014}. }
\label{fig:NMI}
\end{figure*}

We compute the communities at each step using both DyCPM and OLCPM (Communities yielded by DyCPM and OCPM are identical). Then, for each snapshot, we compute the NMI according to \cite{lancichinetti2009b}.
Results are displayed in Figure \ref{fig:NMI}. We show results for k=3 and k=4, which yield the best results.

The average NMI over all snapshots is provided in Table \ref{tab:ANMI}.
\begin{table} [!h] 
    \begin{center}
\begin{tabular}{|c||c|c|c|c|}
        \hline
        \textbf{Algorithm} & DyCPM k=3 & DyCPM k=4 & OLCPM k=3 & OLCPM k=4\\ 
        \hline
       \textbf{ Average NMI} & 0.024	 & 0.004  & 0.059  & 0.044 \\ 
       \hline
\end{tabular} \\  
      
        \caption{Average NMI scores of OLCPM and DyCPM \cite{palla2007} for $k=3$ and $k=4$ on SocioPatterns collaboration networks \cite{fournet2014}.}\label{tab:ANMI}
    \end{center}
\end{table}

We can observe that the average NMI of OLCPM is higher than the original DyCPM, and that values of NMI are also higher for most snapshots.

The longitudinal visualization of Figure \ref{fig:NMI} illustrates the relevance of studying the evolution of a network with a fine granularity: only looking at this plot, we can see that the class structure is not always present in the data. For instance, we can observe that there is no community structure during evenings and weekends, or that the community structure is less observable during several days around lunchtime (Thursday, Friday, second Monday). One can then look in more details to the communities found and their evolution to interpret these observations.
In this example, we were able to run DyCPM because of the small size of the network, the restriction to one-hour interval, and the limitation to 9 days of data, but, as shown previously, it would not be possible to extend this analysis to a much larger number of steps due to the increase in complexity.

\section{Conclusion}

In this paper, we proposed OLCPM framework to discover overlapping and evolving communities in social networks. We proposed  OCPM, an online version of CPM \cite{palla2005}, working on a fully dynamic network model, i.e., described as flows of events, where nodes or edges can be added or removed over time. Instead of calculating all $k$-cliques for the whole network at each event occurring in the network, our method updates only the community structure alongside the node or the edge involved in the event. This local update of the community structure provides a significant improvement in computational time.

To cope with the covering problem of CPM, nodes belonging to OCPM communities are considered as core nodes and we proposed a post-process based on label propagation to discover peripheral nodes.

The experimental results of our framework in both artificial and real-world networks show good performance in both computing time and quality detection.

Our method has some drawbacks, some of which are related to CPM itself, like the dependency of the parameter $k$ (clique size). We intend to propose a heuristic for finding appropriate values of $k$. 

Currently, the post-process is run from scratch at each step, and although it is not as costly as a clique-finding problem, running it at each step for a large network can become very costly. For future research, it would be interesting to extend OLCPM by developing an online version of the post-process.

\section*{References}

\bibliography{mybibfile}

\newpage
\appendix
\appendixpage

This Appendix contains the algorithms defining the OCPM and OLCPM methods.



  
\IncMargin{1em}
\begin{algorithm}[!h]
\SetKwData{Left}{left}
\SetKwData{This}{this}
\SetKwData{Up}{up}
\SetKwFunction{Union}{Union}
\SetKwFunction{FindCompress}{FindCompress}
\SetKwInOut{Input}{input}
\SetKwInOut{Output}{output}
\Input{$K, G, AC, DC, SE$}
\Output{Update $AC, DC, G$}
\BlankLine
\For{$ev \in SE$}{
\Switch{$e$}{
\Case{Add Node }{
$V \leftarrow V \cup\{n\}$\;
break\;
}
\Case{Add Edge}{
$E \leftarrow E \cup\{(i,j)\}$\;
\eIf{ $(C_{i} \neq \varnothing)  or  (C_{j} \neq \varnothing)$ }{
$AddNonExternalEdge(i, j, t, AC, DC, G)$\;
}
{
$AddExtrenalEdge(i, j, t, AC, G)$\;
}
break\;
}
\Case{Remove Node }{
$V \leftarrow V \backslash \{n\}$\;
$E \leftarrow E \backslash \{\forall(i,j), i=n$ or $j=n \}$\;
\If{ $(C_{n} \neq \varnothing) $ }{
$RemoveInternalNode(n, t, AC, DC, G)$\;
}
break\;
}
\Case{Remove Edge }{
$E \leftarrow E \backslash\{(i,j)\}$\;
\If{ $(C_{i} \cap C_{j} \neq \varnothing) $ }{
$RemoveIntrnalEdge(i, j, t, AC, DC, G)$\;
}
break\;
}
}
}
\caption{Online Clique Percollation Method (OCPM)}
\label{algo:OCPM}
\end{algorithm}\DecMargin{1em}
\IncMargin{1em}
\begin{algorithm}[!h]
\SetKwData{Left}{left}
\SetKwData{This}{this}
\SetKwData{Up}{up}
\SetKwFunction{Union}{Union}
\SetKwFunction{FindCompress}{FindCompress}
\SetKwInOut{Input}{input}
\SetKwInOut{Output}{output}
\Input{$i, j, t, AC, DC, G$}
\Output{Update $AC, DC$}
\BlankLine
$KC \leftarrow KCliques(\{i,j\},G)$\;
$KC \prime \leftarrow \{kcl \in KC,\forall cm \in AC, kcl \nsubseteq cm \}$\;
$AKC \leftarrow AdjKCliques(KC\prime)$\;
\For{$c \in AKC$}{
$C_{c} \leftarrow \{ \}$\;
\For{$e \in c $}{
$C_{c} \leftarrow C_{c} \cup \{ C_{e}\}$\;
}
$AdjC \leftarrow \{ \}$\;
\For{$cm \in C_{c}$}{ 
\If{$\vert cm \cap c \vert \geqslant K-1 $}{
$AdjC \leftarrow AdjC\cup\{cm\}$\;
}
}
\eIf{$AdjC\neq\varnothing$}{
\For{$cm \in AdjC$}{ 
$Growth(cm, c, AC)$\;
}
\If{$\vert AdjC\vert > 1$}{
$Merge(AdjC, t, AC, DC)$\;
}
}
{
$Birth(c, t, AC)$\;
}

}
\caption{Add Non-External Edge}
\label{Add_internal_edge}
\end{algorithm}\DecMargin{1em}
\IncMargin{1em}
\begin{algorithm}[!h]
\SetKwData{Left}{left}
\SetKwData{This}{this}
\SetKwData{Up}{up}
\SetKwFunction{Union}{Union}
\SetKwFunction{FindCompress}{FindCompress}
\SetKwInOut{Input}{input}
\SetKwInOut{Output}{output}
\Input{$i, j, t, AC, G$}
\Output{Update $AC$}
\BlankLine
$KC \leftarrow KCliques(\{i,j\},G)$\;
$AKC \leftarrow AdjKCliques(KC)$\;
\For{$c \in AKC$}{
$Birth(c, t, AC)$\;
}
\caption{Add External Edge}
\label{Algo:AddExternalEdge}
\end{algorithm}\DecMargin{1em}
\IncMargin{1em}
\begin{algorithm}[!h]
\SetKwData{Left}{left}
\SetKwData{This}{this}
\SetKwData{Up}{up}
\SetKwFunction{Union}{Union}
\SetKwFunction{FindCompress}{FindCompress}
\SetKwInOut{Input}{input}
\SetKwInOut{Output}{output}
\Input{$n, t, AC, DC, G$}
\Output{Update $AC, DC$}
\BlankLine
\For{$c \in C_{n}$}{
$KC \leftarrow KCliques(c,G)$\;
\eIf{$KC = \varnothing$}{
$Death(c, t, AC, DC)$\;
}
{
$Shrink(c,n, AC)$\;
$Split(c, t, AC, G)$\;
}
}
\caption{Remove Internal Node}
\label{Remove_internal_node}
\end{algorithm}\DecMargin{1em}
\IncMargin{1em}
\begin{algorithm}[!h]
\SetKwData{Left}{left}
\SetKwData{This}{this}
\SetKwData{Up}{up}
\SetKwFunction{Union}{Union}
\SetKwFunction{FindCompress}{FindCompress}
\SetKwInOut{Input}{input}
\SetKwInOut{Output}{output}
\Input{$i, j, t, AC, DC, G$}
\Output{Update $AC, DC$}
\BlankLine
\For{$c \in C_{i j}=\{\forall cm, cm \in (C_{i}\cap C_{j})\}$}{ 
$KC \leftarrow KCliques(c,G)$\;
\eIf{$KC = \varnothing$}{
$Death(c, t, AC, DC)$\;
}
{
$Split(c, t, AC, G)$\;
}
}
\caption{Remove Internal edge}
\label{Remove_internal_edge}
\end{algorithm}\DecMargin{1em}
\IncMargin{1em}
\begin{algorithm} [!h]
\SetKwData{Left}{left}
\SetKwData{This}{this}
\SetKwData{Up}{up}
\SetKwFunction{Union}{Union}
\SetKwFunction{FindCompress}{FindCompress}
\SetKwInOut{Input}{input}
\SetKwInOut{Output}{output}
\Input{$ SN$:Set of nodes,$G$}
\Output{$ SKC$: Set of Set of nodes}
\BlankLine
\For{$n \in SN$}{
$N \leftarrow Neighbors(n,G)$\;
$L \leftarrow L\cup \{N\}$\;
}
$CN \leftarrow \cap_{l\in L} l$\;
$SCL \leftarrow MaximalCliques(CN, K)$\;
\caption{KCliques}
\label{KCliques}
\end{algorithm}\DecMargin{1em}
\IncMargin{1em}
\begin{algorithm} [!h]
\SetKwData{Left}{left}
\SetKwData{This}{this}
\SetKwData{Up}{up}
\SetKwFunction{Union}{Union}
\SetKwFunction{FindCompress}{FindCompress}
\SetKwInOut{Input}{input}
\SetKwInOut{Output}{output}
\Input{$ Adjc, t, AC, DC$}
\Output{Update$  AC,DC$}
\BlankLine
$mc \leftarrow c, |c|= max_{x \in Adjc} |x| $\;
$Adjc \leftarrow Adjc \backslash \{mc\}$\;
$mc \leftarrow \cup_{x \in Adjc}$ $x$\;
\For{$x \in Adjc$}{
$Death(c, t, AC, DC)$\;
}
\caption{Merge}
\label{Merge}
\end{algorithm}\DecMargin{1em}

\IncMargin{1em}
\begin{algorithm}[!h]
\SetKwData{Left}{left}
\SetKwData{This}{this}
\SetKwData{Up}{up}
\SetKwFunction{Union}{Union}
\SetKwFunction{FindCompress}{FindCompress}
\SetKwInOut{Input}{input}
\SetKwInOut{Output}{output}
\Input{$ c, t, AC, G$}
\Output{Update$  AC$}
\BlankLine
$KCc \leftarrow KCliques(c, G)$\;
$Adjc \leftarrow AKCliques(KCc)$\;
$c \leftarrow mc, |mc|= max_{x \in Adjc} |x| $\;
$Adjc \leftarrow Adjc \backslash \{c\}$\;
\For{$cm \in Adjc$}{
$Birth(cm, t, AC)$\;
}
\caption{Split}
\label{Split}
\end{algorithm}\DecMargin{1em}

\IncMargin{1em}
\begin{algorithm} [!h] 
\SetKwData{Left}{left}
\SetKwData{This}{this}
\SetKwData{Up}{up}
\SetKwFunction{Union}{Union}
\SetKwFunction{FindCompress}{FindCompress}
\SetKwInOut{Input}{input}
\SetKwInOut{Output}{output}
\Input{$ AC, G$}
\Output{Update$  AC$}
\BlankLine
$PN \leftarrow \{ n, n \in c \forall c\in AC \}$\;
//Label spreading\\

\For{$c \in AC$}
{
$d \leftarrow 1$\;
$S \leftarrow c$\;
$\textbf{x:} N \leftarrow Nieghbors(S)$\;
$N \leftarrow N \cap PN$\;
\If{$N \neq\ \varnothing$}
{
\For{$n \in N$}
{
$Label(n,c.id,d)$\;
}
$S \leftarrow N$\;
$d \leftarrow d+1$\;
\textbf{goto x:}
}
}
//label Analyses\\

\For{$n \in PN$}{
$idc \leftarrow LabelAnalysis(n.label)$\;
$Growth(idc, n)$\;
}
\caption{OLCPM}
\label{algo:OLCPM}

\end{algorithm}\DecMargin{1em}


\end{document}